\documentclass[traditabstract]{aa}
\usepackage{graphicx}
\usepackage{natbib}
\usepackage{txfonts}
\bibpunct{(}{)}{;}{a}{}{,}
\begin{document}

  \title{The red giant branch phase transition: Implications for the RGB luminosity function bump and detections of Li-rich red clump stars}
  
  \author{Santi Cassisi\inst{1,2}, Maurizio Salaris\inst{3}, \and  Adriano Pietrinferni\inst{1}}

\institute{INAF~$-$~Osservatorio Astronomico di Teramo, Via M. Maggini, I$-$64100 Teramo, Italy, 
            \email{cassisi@oa-teramo.inaf.it}  
            \and
            Instituto de Astrof{\'i}sica de Canarias, Calle Via Lactea s/n, E-38205 La Laguna, Tenerife, Spain
            \and
            Astrophysics Research Institute, 
           Liverpool John Moores University, 
           IC2, Liverpool Science Park, 
           146 Brownlow Hill, 
           Liverpool L3 5RF, UK, \email{M.Salaris@ljmu.ac.uk} 
           }

 \abstract{We performed a detailed study of the evolution of the luminosity of He-ignition stage and of the 
red giant branch bump luminosity during the red giant branch phase transition for various metallicities.  
To this purpose we calculated a grid of stellar models that sample the mass range of the transition 
with a fine mass step equal to ${\rm 0.01M_\odot}$. 
We find that for a stellar population with a given initial chemical composition, there is 
a critical age (of 1.1-1.2~Gyr)
around which a decrease in age of just 20-30 million years causes a drastic drop in the red giant branch tip brightness. 
We also find a narrow age range (a few $10^7$ yr) around the transition, characterized by the luminosity of the  
red giant branch bump being brighter than the luminosity of He ignition.  
We discuss a possible link between this occurrence and observations of Li-rich core He-burning stars.
}
\authorrunning{S. Cassisi et al.}
\titlerunning{Fine details of the RGB phase transition}
  \maketitle


\section{Introduction}

Theoretical stellar evolution calculations show that,  
for any given chemical composition, there is a minimum initial mass -- usually named ${\rm M_{HeF}}$ -- that 
is able to ignite He-burning at the end of the red giant branch (RGB) phase within a He core not affected by electron degeneracy. 
Below this threshold,  
the RGB phase lasts longer and He-burning ignition occurs via a mildly violent He flash in an electron degenerate core. 
After the detailed analysis by \cite{sgr:90}, it is customary to use the designation {\sl RGB phase transition} to describe the changes that occur in both the morphology of the 
colour-magnitude-diagram and the integrated spectral energy distribution of a stellar population, when stars with mass lower than ${\rm M_{HeF}}$ begin to populate  the RGB.

All stars with initial masses below  ${\rm M_{HeF}}$ (that we denote as {\sl \emph{low-mass stars}}), owing to the large -- and similar --  level of electron degeneracy in their  
cores, reach a similar value of the He-core mass (${\rm M_{cHe}}$) at He ignition. Owing to 
the ${\rm M_{cHe}}$-luminosity relation for electron degenerate cores, these stars will also attain a similar   
bolometric luminosity at He ignition, which coincides with the brightest point along the RGB (this latter denoted as RGB tip). 
When the initial mass reaches values around  ${\rm M_{HeF}}$, there is a narrow mass range 
(${\rm\sim0.15M_\odot}$; almost independent of chemical composition) populated by 
objects that do not develop a significant level of electron degeneracy in the He core and that attain 
He ignition via a very mild He flash. In this regime ${\rm M_{cHe}}$ at ignition decreases drastically with increasing stellar mass. 
Beyond this range, when the initial mass is higher than ${\rm M_{HeF}}$, the value of ${\rm M_{cHe}}$ 
at ignition increases almost linearly with mass
as a consequence of the increasing mass of the convective core during the previous central H-burning stage.

Given that both luminosity and lifetimes of the following evolutionary phases 
depend on the value of ${\rm M_{cHe}}$ at He ignition, 
the discontinuity of the trend of the He-core mass with total mass around ${\rm M_{HeF}}$ does 
affect the properties of core and shell He-burning stages.

At the same time, the differential luminosity function (LF -- star counts as a function of magnitude) of the RGB of 
old stellar populations, such as globular clusters, displays a characteristic 
local maximum in the star counts, commonly denoted as the {\sl \emph{RGB bump}} \citep{thomas:67,iben:68, king:85}. 
The luminosity of the RGB bump corresponds to the stage when 
the H-burning shell crosses the H-abundance discontinuity left over by the outer convection zone at the completion of the first dredge up 
during the early RGB evolution. 
During this crossing, the luminosity of the star drops temporarily, before starting to increase again after the H shell 
has moved beyond the discontinuity. There is therefore a luminosity range along the RGB that is crossed three times, 
and this causes the appearance of the bump in the LF. 
The RGB bump brightness is an important tracer of the interior chemical stratification of low-mass 
RGB stars, and it has been at the crossroad of several theoretical and
observational investigations \citep[see][and references therein]{cs:97, cassisi:02, scw02, nataf:13, nataf:14, joergen:15}.

For a 
fixed initial stellar mass, the RGB bump becomes brighter when the initial He abundance increases and/or the metallicity decreases, 
whilst for a given chemical composition, the bump becomes brighter when the initial mass increases.
These trends are explained by the varying depth of the convective envelope at the completion of the dredge up. 
The shallower (in terms of distance from the centre in mass units) 
the convective envelope at the dredge up, the higher the mass of the He-core 
when the H-burning shell encounters the H-discontinuity, and the brighter the RGB bump.
In stars with mass above ${\rm M_{HeF}}$, 
the thermal conditions required for He ignition are achieved earlier, as a consequence of the vanishing level of electron degeneracy, so that 
triple-$\alpha$ nuclear burning starts before the advancing H-burning shell encounters the H-discontinuity, and the RGB bump disappears.

In this paper we investigate the occurrence of the RGB bump in the thus far unexplored mass range that corresponds to the RGB phase transition.
Section~2 describes our models, while Sect.~3 presents the results concerning 
the RGB bump. Section~4 discusses a possible connection with the presence of lithium-rich red clump stars.

\section{The theoretical models}

All stellar models presented here have been computed with the same code and physics inputs as are employed for the 
BaSTI\footnote{http://www.oa-teramo.inaf.it/BASTI} stellar model library \citep[][]{basti, basti:06}, 
along with a scaled-solar heavy element mixture \citep[see][for more details]{basti}. We computed models with 
convective core overshooting during the main sequence by using the same prescriptions 
adopted in \cite{basti}, and we mapped the 
whole RGB phase transition with a very fine mass step of ${\rm 0.01M_\odot}$. This choice allows us to trace any change in the evolutionary 
properties of stars along the transition, which would probably be missed when adopting a coarser mass grid.
Models have been computed for metallicities  Z=0.004 (Y=0.251 -- mass between 1.75 and 2.00${\rm M_\odot}$),  
Z=0.008 (Y=0.256 -- mass between 1.80 and 2.00${\rm M_\odot}$), and Z=0.0198 (Y=0.2734 -- mass between 1.99 and 2.05${\rm M_\odot}$).  
These chemical compositions correspond to grid points in the original BaSTI archive, and our calculations 
effectively represent an extension of the BaSTI model library.

\begin{figure}
\centering
\includegraphics[scale=.470]{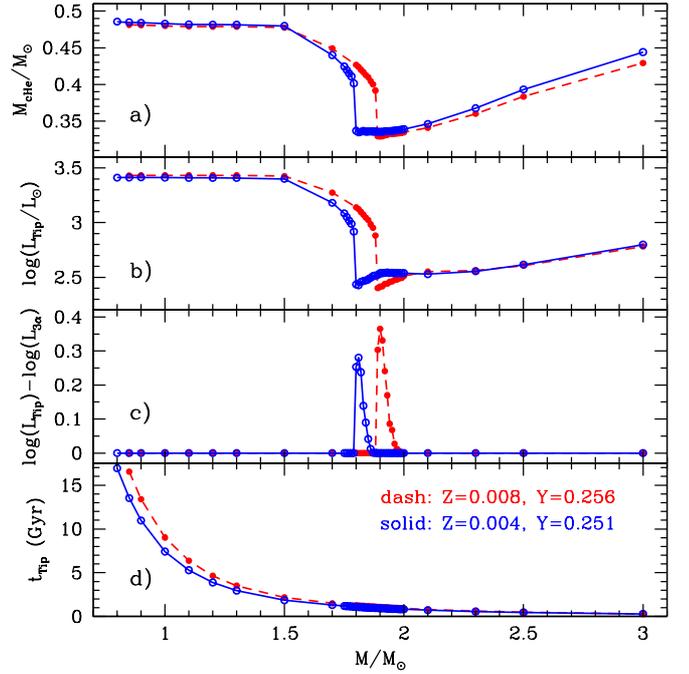}
\vskip -0.5cm
\caption{Selected properties at the RGB tip as a function of the initial total mass for the two labelled chemical compositions: {\sl panel a)} - 
the He-core mass; {\sl panel b)} - bolometric luminosity; {\sl panel c)} - difference in bolometric luminosity between the 
RGB tip (${\rm L_{Tip}}$) and He-ignition; {\sl panel d)} - age .}
\label{fig:tip}
\end{figure}

Figure~\ref{fig:tip} displays selected evolutionary properties at the brightest point along the RGB evolution of our models 
for Z=0.004 and z=0.008. The models shown in this figure were computed including  
convective core overshooting during the central H-burning stage, as in BaSTI calculations.     
The convective cores have been extended beyond the 
boundary fixed by the Schwarzschild criterion by 
an amount $\lambda_{ov} H_p$, where $H_p$ is the pressure scale height at the Schwarzschild border and 
$\lambda_{ov}$ a free parameter.  
For masses higher than or equal to 1.7$M_{\odot}$, the models employ $\lambda_{ov}$=0.20$H_p$; for stars less massive than 
1.1$M_{\odot}$ $\lambda_{ov}$=0, while in the intermediate range 
of the model grid (M=1.1, 1.2, 1.3, 1.4, 1.5, and 1.6$M_{\odot}$), $\lambda_{ov}$
varies as $\lambda_{ov}$=($M/M_{\odot}$-0.9)/4 \citep[see][for details]{basti}.
We did not account for any overshooting from the bottom of the convective envelope. 

The RGB transition can be seen at masses around ${\rm 1.8-1.9M_\odot}$ depending on the metallicity. 
The core mass ${\rm M_{cHe}}$ at He ignition and the RGB tip luminosities drop sharply at ${\rm M=1.80M_\odot}$ for Z=0.004 and ${\rm M=1.89M_\odot}$ for Z=0.008, 
because of the drastic decrease in the level of electron degeneracy in the He core.  
As shown by panel c) of Fig.~\ref{fig:tip}, this drop in the RGB tip luminosity coincides with an unexpected feature. 
Considering, for example, the Z=0.004 models, He ignition occurs at  
luminosities (also significantly) fainter than the RGB tip for all masses between $\sim$1.79
and ${\rm \sim1.85M_\odot}$. The same behaviour can be seen for the Z=0.008 models, although shifted to slightly higher masses, 
as expected by the increase in the transition mass ${\rm M_{HeF}}$ with metallicity \citep[see][and references therein]{cc:93}.
Outside this narrow mass range, the RGB tip and He ignition do coincide, as usual.

The mass range under scrutiny is so narrow that 
a very fine mass spacing (${\rm \Delta{M}\sim 0.01M_\odot}$) is required to reveal this behaviour. This is why, to the best of our knowledge, it has not been detected in any previous investigation. 
In the following section, we show how this is related to the combined occurrence of He-burning ignition 
and the H-burning shell crossing of the H-abundance discontinuity left over at the first dredge up.

\section{The RGB bump luminosity at the RGB phase transition}

\begin{figure}
\centering
\includegraphics[scale=.460]{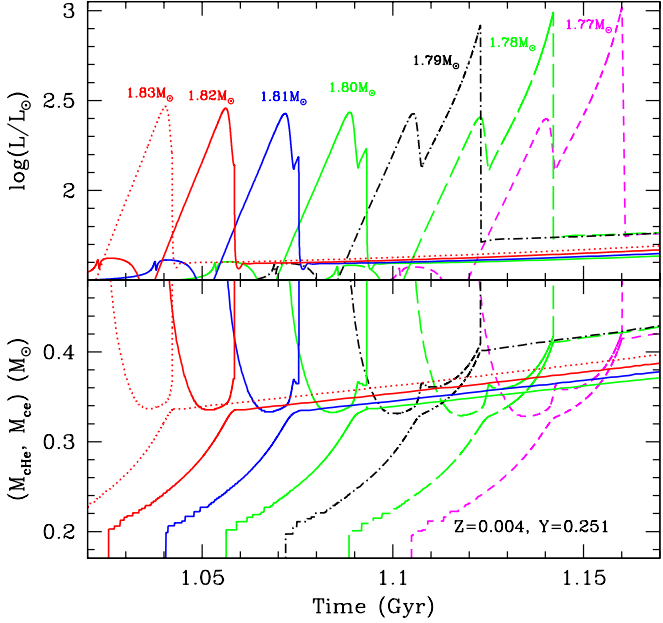}
\vskip -0.7cm
\caption{{\sl Top panel}: bolometric luminosity as a function of time during the RGB evolution of selected masses  
(see labels) for Z=0.004 and Y=0.251. {\sl Bottom panel}: as in the top panel but for the mass location of the boundary of the He core and at the 
bottom of the convective envelope.}
\label{fig:ltime}
\end{figure}

Figure~\ref{fig:ltime} shows the evolution of the bolometric luminosity as a function of time during the RGB 
for models with Z=0.004, when sampling the mass range around the minimum of the RGB tip luminosity as shown in Fig.~\ref{fig:tip}. 
For masses up to ${\rm 1.79M_\odot}$ and with increasing age, the luminosity displays a local maximum and a temporary decrease 
(the RGB bump) followed by a steep increase 
up to a maximum value (RGB tip) before a sharp drop after He ignition in the electron degenerate core.
This is the standard evolution, with the RGB bump fainter than the tip.
 
When increasing the mass by only ${\rm 0.01M_\odot}$, the 
evolution changes drastically. The ${\rm 1.80M_\odot}$ track clearly
displays the RGB bump, with the luminosity dropping to a 
minimum value consistent with the value attained by lower mass models. However, soon after the luminosity starts 
increasing again (following the standard behaviour) He-burning ignites in the core at a luminosity 
fainter than the maximum pre-bump value. 
As a consequence, for this model, which corresponds to the mass with the largest brightness difference between RGB tip and He ignition in panel c) of 
Fig.~\ref{fig:tip}, the RGB tip does {\sl not} mark the beginning of core He-burning, but instead corresponds to the onset of the 
RGB bump phase. 

The lower panel of Fig.~\ref{fig:ltime} displays the variation with time of both the location (in mass units) of the 
bottom of the convective envelope and the boundary of the He core, for the same 
models shown in the top panel. In the ${\rm 1.80M_\odot}$ model, the H-burning shell encounters
the H-abundance discontinuity (at the time when the He-core boundary reaches the layer corresponding to the bottom of the convective  
envelope at its maximum extension) a short time before the triple-$\alpha$ ignition in the 
core. The signature of He-burning ignition -- via an extremely mild He-flash -- is denoted by the small bump (at t$\sim$1.09~Gyr) 
in the trend of the bottom of 
the convective envelope with time, owing to the energy flux released at He ignition.

The ${\rm 1.81M_\odot}$ displays the same behaviour, but He ignition occurs even earlier,  
right after the minimum in luminosity of the RGB bump phase. 
With increasing initial mass, the He ignition occurs progressively earlier, before the 
luminosity reaches the minimum during the RGB bump phase, and manifests itself as a kink during the 
decrease in the luminosity due to the crossing of the H-abundance 
discontinuity. This kink eventually disappears\footnote{This kink can be barely seen in the ${\rm 1.82M_\odot}$ at ${\rm\log(L/L_\odot)\approx 2.142}$.}, 
and for masses greater than 
${\sim 1.84M_\odot}$, He ignition occurs before the H-burning shell encounters the H-abundance discontinuity, so that once again the absolute maximum in 
luminosity along the RGB marks the start of core He-burning stage.

\begin{figure}
\centering
\includegraphics[scale=.460]{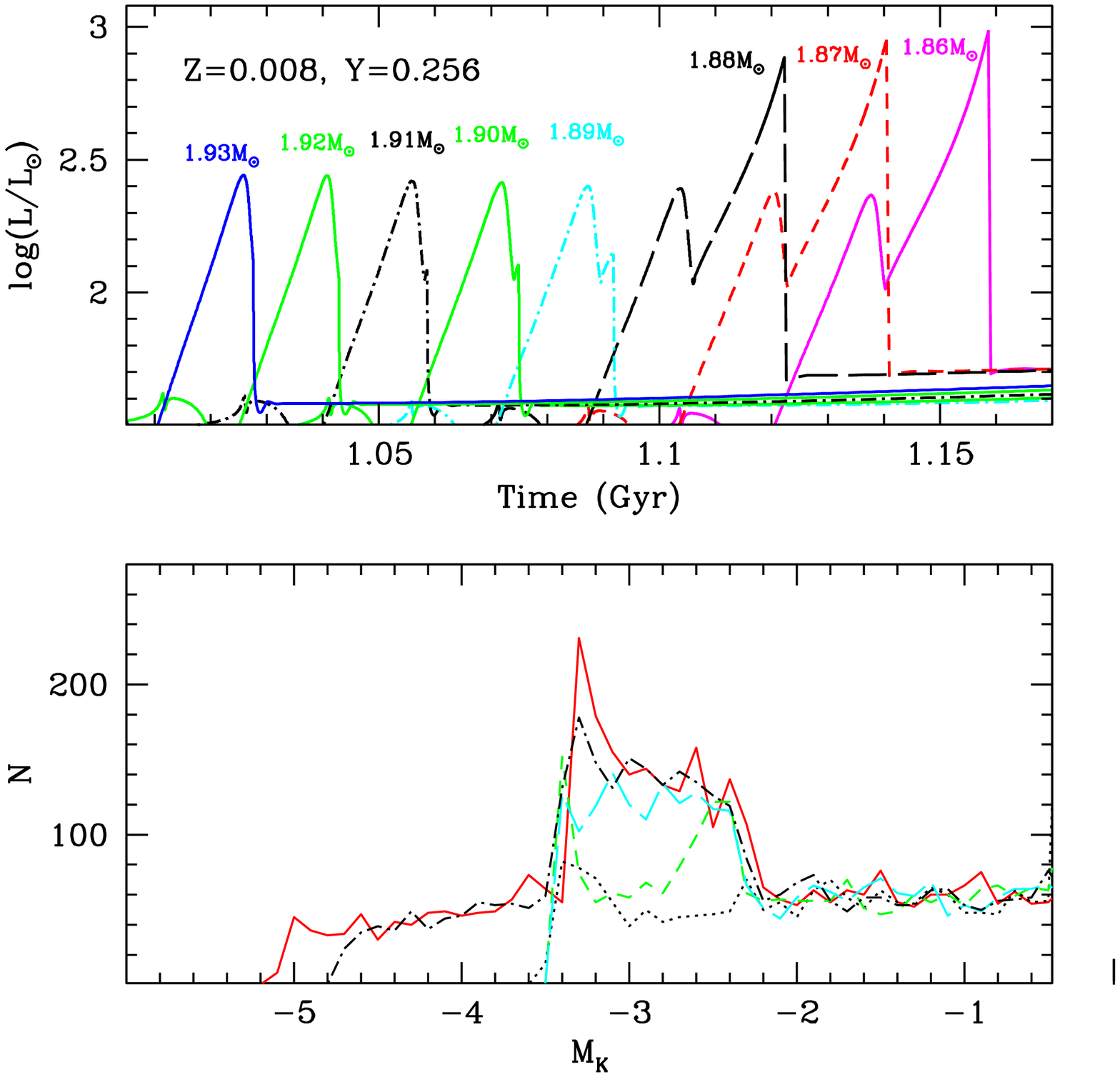}
\vskip -0.2cm
\caption{{\sl Top panel}:  as top panel of fig.~\ref{fig:ltime} but for the chemical composition Z=0.008, Y=0.256. {\sl Bottom panel}: 
LF (0.1~mag bin size) of the upper RGB of $\sim 10^6 {\rm M_{\odot}}$ clusters for the same chemical composition, and ages t=1.14 
(solid line), 1.12 (dotted), 1.10 (short dashed), 1.08 (long dashed), 1.06 (dot-dashed) Gyr, respectively.}
\label{fig:lzz}
\end{figure}

This behaviour is common to all chemical compositions covered by our analysis. 
The upper panel of Fig.~\ref{fig:lzz}  
displays, for example, the trend of the surface luminosity 
as a function of time during the RGB evolution for models at the RGB transition and initial composition with 
Z=0.008 and Y=0.256. 
The only difference with Z=0.004 models is a shift to higher masses, owing to 
the increase in the transition mass ${\rm M_{HeF}}$ with Z.

The sharp change in luminosity at He ignition implies that for stellar populations of a given chemical composition 
there is a critical age equal to $\sim$1.1~Gyr at Z=0.004 and $\sim1.15$~Gyr at solar chemical composition, 
around which a decrease in age of just $2-3\times10^7$~yr causes a drastic drop in the RGB tip brightness 
(by $\sim1.5$~mag). 
This can be seen clearly in the lower panel of Fig.~\ref{fig:lzz}, which displays the $K$-band (that traces 
well the bolometric luminosity of RGB stars) LF 
of $\sim 10^6 {\rm M_{\odot}}$ \citep[assuming a][initial mass function]{kroupa}  
synthetic clusters with ages equal to 1.14, 1.12, 1.10, 1.08, and 1.06~Gyr (Z=0.008, Y-0.256), 
as simulated with our models. 
The brightness of the RGB tip decreases abruptly by $\sim$1.5~mag at ages between 1.12 and 1.10~Gyr. 
At an age of 1.10~Gyr (when the RGB tip coincides with the RGB bump), 
the number of stars within about one magnitude of the RGB tip is about twice that for the two older ages. 
This {\sl enhanced} number of stars near the RGB tip disappears when moving to ages equal to 1.08 and 1.06~Gyr, owing to the 
reduced evolutionary lifetimes of the evolving RGB mass.

The behaviour of the RGB LF around the transition is very difficult to observe. One needs a sample of clusters in the 
relevant narrow age range discussed here, which is massive enough to have a well-populated upper RGB 
(we estimate that a total mass of at least $\sim 5\times 10^{5} {\rm M_{\odot}}$ is needed) that allow to detect unanbiguously RGB tip and bump, 
and all approximately at the same 
distance, or at least with reasonably well determined relative distances.

\section{A link with Li-rich red clump stars?}

Lithium is a fragile element that is easily destroyed by $\alpha$ captures at temperatures 
higher than $\sim 2.5\times10^6$~K. During the early RGB evolution, 
Li in the envelope is exposed to the high temperatures of the interiors with a consequent dilution because
of the deepening of the outer convection zone. Starting from the present   
interstellar medium abundance ${\rm A(Li)\sim3.3}$~dex\footnote{As is customary, we define ${\rm A(Li)=\log[n(Li)/n(H)]+12}$.} 
\cite{gs:98}, a post dredge-up A(Li)$\sim1.5$~dex is expected in Population I stars, 
the exact value depending on the stellar mass and initial chemical composition \citep[][]{cb:00}. 
Spectroscopic surveys \citep[see, i.e.,][]{brown:89} have, however, discovered a small fraction ($\sim$1\%) 
of RGB stars with Li abundances that exceed this value -- hence named {\sl \emph{Li-rich}} stars -- 
and indeed they pose a challenge for standard stellar evolution theory. 

Several scenarios have been envisaged to explain the occurrence of 
Li-rich giants \citep[see, i.e.,][and references therein]{sb:99,cb:00,dw:00,d:12}. 
For RGB stars brighter than the bump, the general picture \citep[][]{sb:99, cb:00} is that a phase may exist during which 
newly synthesized Li can be circulated by convection in the outer convective envelope, resulting in a short phase of Li richness.
Briefly, the mean molecular weight discontinuity left over by the first dredge-up is expected to strongly inhibit any extra-mixing below 
the convective 
envelope. However, when this discontinuity is erased after the RGB bump, extra-mixing processes are possible, as shown by the evolution of the 
${\rm ^{12}C/^{13}C}$ ratio in field giants at various metallicities
\citep[see][and references therein]{cb:00}. 
The ${\rm ^3He}$ produced through the $pp$-chain on the main sequence and engulfed by the convective envelope during the dredge-up 
should then first be transported down in the vicinity of the H-burning shell 
at temperatures high enough for burning through the ${\rm ^3He}$($\alpha$, $\gamma$)${\rm ^7Be}$ reaction. 
Then, the produced ${\rm ^7Be}$ should be circulated up to convective layers cool enough where it can decay into ${\rm ^7Li}$ by 
${\rm ^7Be}$(${\rm e^-}$, $\nu$)${\rm ^7Li,}$ without ${\rm ^7Li}$ undergoing proton captures.
Whether and for how long stars appear Li-rich depends critically on the mass flow rate in the extra-mixing region \citep[][]{sb:99}.
It is generally expected \citep[e.g.,][]{cb:00} that when climbing the RGB beyond the bump, 
the extra mixing must reach even hotter layers to be able to modify the surface ${\rm ^{12}C/^{13}C,}$ as observed, with the net effect 
of Li destruction. Therefore, the RGB Li-rich phase turns out naturally to be a short-lived phase, close to the bump luminosity, as generally observed.  
\citep[But see][for observations of Li-rich RGB stars that are much brighter than the bump.]{monaco:11}

\begin{figure}
\centering
\includegraphics[scale=.460]{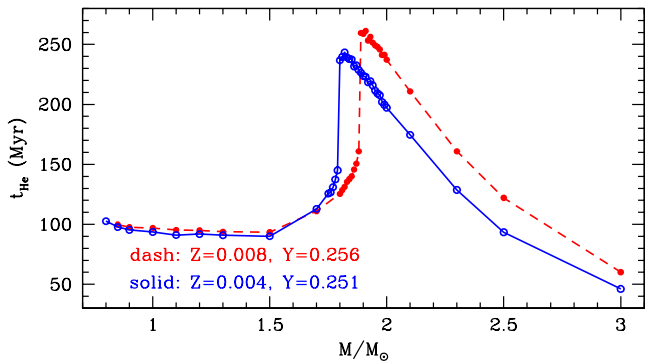}
\vskip -3.8cm
\caption{Lifetime during the central He-burning stage for selected stellar masses along the RGB phase transition, with the labeled chemical 
compositions.}
\label{fig:the}
\end{figure}

The discovery of Li-rich RC stars \citep[][]{kumar:11, silva:14} challenges the scenario described above, because RGB stars need to  
maintain high levels of Li enrichment from the RGB bump until He ignition. 
Although it is possible that surface Li enrichment is a consequence of extra-mixing processes during the He flash 
\citep[see][for a discussion on this possibility]{silva:14}, it is difficult to reconcile this occurrence with 
the relative paucity of Li-rich, core He-burning stars.
Here we just suggest a working hypothesis to explain at least some RC Li-rich stars, 
to be further investigated with a detailed analysis. 

We have seen that at the RGB phase transition there is a narrow mass range that ignites He 
shortly after the RGB bump (timescale of the order of $10^6$~yr or less). 
If extra-mixing processes can efficiently contribute to quickly increasing the surface Li 
abundance at the bump, Li production is very likely  
not followed by any depletion (and change in the ${\rm ^{12}C/^{13}C}$  ratio) before He ignition is achieved, 
so the star should appear as a Li-rich RC object.
An important observational consequence is that this class of Li-rich RC stars 
should show a carbon isotopic ratio similar to the value expected after the first dredge-up.
This RGB mass range is extremely narrow, so that the possibility of actually observing the RC progeny of these stars should be extremely small. 
However, as shown in Fig.~\ref{fig:the}, models with initial mass around the RGB transition display the longest core He-burning lifetime, 
as a consequence of the lower He-core mass, hence fainter RC luminosity \citep[see also][for a detailed discussion of this issue]{cs13}.
This would somewhat increase the probability of observing the RC progeny of RGB transition stars amongst the field disk population. 

The RC Li-rich star recently identified by \cite{silva:14} cannot be associated to this potential formation channel, given 
the lower ${\rm ^{12}C/^{13}C}$ compared to the expected post dredge-up value\footnote{Also the estimated mass 
M=${\rm 1.536^{+0.059}_{-0.056}M_\odot}$ is $\sim 0.4~{\rm M_\odot}$ 
lower than the initial mass at the RGB transition for about half-solar metallicity. However, 
this could potentially be explained by efficient RGB mass loss.}. However, there are Li-rich objects that appear to be RC stars 
from their location in the Hertzsprung-Russell diagram and that have a normal post dredge-up ${\rm ^{12}C/^{13}C}$ ratio \citep{kumar:11}. 
A more precise estimate of their luminosity and $T_{eff}$ may allow us to better assess both evolutionary status and mass, so as 
to establish whether they can be related to the progeny of RGB transition objects.

\begin{acknowledgements}

SC warmly acknowledges financial support by PRIN-INAF2014 (PI: S. Cassisi) and by the Economy and Competitiveness Ministry of the Kingdom of Spain (grant AYA2013-42781-P). 
AP acknowledges financial support by PRIN-INAF2012 (PI: L. Bedin).
We thank Laura Greggio for having suggested the calculations that led to these results.
\end{acknowledgements}

\bibliographystyle{aa}
\bibliography{bumpers}

\end{document}